\documentclass{jpp}
\usepackage{graphicx} 
\usepackage{epstopdf}
\usepackage{amssymb}
\usepackage{amsmath} 
\DeclareMathOperator{\sech}{sech}
\usepackage{float} 
\usepackage{wrapfig}
\usepackage{subcaption}
\usepackage{natbib} 
\usepackage[dvipsnames]{xcolor}

\usepackage[hidelinks]{hyperref} 
\usepackage[all]{hypcap} 
\title{Fluid Simulations of Three-Dimensional Reconnection that Capture the Lower-Hybrid Drift Instability}

\author{F. Allmann-Rahn\aff{1} \corresp{\email{far@tp1.rub.de}}, S.~Lautenbach\aff{1}, R.~Grauer\aff{1}, \and R.~D.~Sydora\aff{2}}

\affiliation{
\aff{1}Institute for Theoretical Physics I, Ruhr University Bochum, Germany
\aff{2}Department of Physics, University of Alberta, Edmonton, Alberta T6G 2E1, Canada

}

\begin{document}

\maketitle

\begin{abstract}

Fluid models that approximate kinetic effects have received
attention recently in the modelling of large scale plasmas
such as planetary magnetospheres. In three-dimensional reconnection,
both reconnection itself and current sheet
instabilities need to be represented appropriately.
We show that a heat flux closure based on pressure gradients
enables a ten moment fluid model to capture key properties of
the lower-hybrid drift instability (LHDI) within a reconnection simulation.
Characteristics of the instability are examined with kinetic and fluid
continuum models, and its role in the three-dimensional
reconnection simulation is analysed. The saturation level of
the electromagnetic LHDI is higher than expected which leads
to strong kinking of the current sheet. Therefore,
the magnitude of the initial perturbation has significant impact
on the resulting turbulence.

\end{abstract}

\section{Introduction}\label{sec:introduction}

The reconnection of magnetic field lines in a plasma causes release
of magnetic energy. Magnetic reconnection takes place, for example, in
solar flares and the Earth's magnetosphere and thus influences space weather. 
In fusion devices it can induce
further instabilities and affect confinement of the fusion plasma.

In the last decades, reconnection has been the subject of many
numerical studies, especially starting with the GEM challenge
\citep{birn-drake-shay-etal:2001} where a widely used
setup for reconnection simulations was defined. The deeper
understanding of fast reconnection gained through numerical
simulations has been complemented by laboratory experiments
such as the Magnetic Reconnection Experiment (MRX)
\citep{yamada-ji-etal:1997}. In 2015 NASA started the
Magnetospheric Multiscale Mission (MMS) which was the first
spacecraft to directly measure reconnection in the magnetosphere. 

The increase in computational power and the availability of data
from MMS measurements open up new possibilities and define
the directions of current reconnection research. We will briefly
list what we consider to be central research subjects. One important
point is of course the interpretation of MMS measurements, and
numerical simulations are a valuable tool in this regard.
On the other hand, the MMS measurements can
also be used to validate the numerical models. Recent studies showed
agreement with MMS data for kinetic particle-in-cell (PIC)
simulations (e.g.~\citet{nakamura-genestreti-etal:2018}) and
for ten moment fluid simulations \citep{tenbarge-ng-etal:2019}.
Both models retain the anisotropic pressure tensor which
is one mechanism to break the frozen-in condition
in collisionless reconnection \citep{egedal-ng-etal:2019}. Generally,
there is hope for more insights from the MMS mission concerning dissipation
processes (e.g.~Landau damping) that break the frozen-in law.

Another question is that of
three-dimensionality. Due to limited computational resources,
past research focused on two-dimensional reconnection where the
third dimension is assumed to be homogeneous. Under which conditions
this is a good approximation and how much influence
current sheet instabilities and resulting turbulence have,
is still a largely open question. Relevant instabilities
are for example the lower-hybrid drift instability (LHDI) and the
firehose instability \citep{le-stanier-etal:2019} as well
as the current sheet shear instability (CSSI) and Kelvin-Helmholtz/kink
type instabilities \citep{fujimoto-sydora:2012,fujimoto-sydora:2017}.
The LHDI, whose fluid representation is discussed in this work, has been
extensively studied theoretically and in kinetic simulations
(e.g.~\citet{daughton:2003}; \citet{innocenti-norgren-etal:2016}) along with frequent
measurements of LHDI fluctuations being made at magnetospheric reconnection sites,
such as the magnetopause, and in laboratory reconnection experiments.
The free energy source for the instability are inhomogeneities in the magnetic
field and plasma pressure which drive the relative drifts of electrons and ions.
Even though the LHDI does not significantly alter the reconnection rate
it does lead to enhanced anomalous plasma transport which relaxes gradients,
for instance the density, which in turn can give rise to
secondary instabilities such as the CSSI that is driven
by the electron flow shear \citep{fujimoto-sydora:2017}. 
The electromagnetic branch of the LHDI
causes kinking of the current sheet. At later times kink type
instabilities can be induced by the LHDI \citep{lapenta-brackbill-etal:2003}.

Finally, the impact of reconnection on macroscopic systems like
planetary magnetospheres is of great interest. Generally, being
able to simulate large scale, global systems with models more
accurate than MHD brings new opportunities: both for getting
a better understanding of physical processes in the magnetosphere
and for the application in space weather forecast.
Treating global systems with fully kinetic particle-in-cell (PIC)
models is difficult due to their high computational expense.
Therefore, recent studies utilised ten moment fluid models
with good success. \citet{wang-germaschewski-etal:2018}
modelled the magnetosphere of the Jupiter moon Ganymede,
and \citet{dong-wang-etal:2019} studied the interaction
between the solar wind and Mercury's magnetosphere. The
applicability of multi-fluid models for very large scales
also profits from their ability to under-resolve scales and
still yield appropriate results \citep{wang-hakim-etal:2020}.
However, the ten moment model used in aforementioned studies
(first presented in \citet{wang-hakim-etal:2015}) has
some drawbacks, indicating that the approximation
of kinetic effects in the model needs to be improved.

Different directions can be taken concerning the integration
of kinetic effects in fluid models, and we want to highlight
some of them. Kinetic effects include non-collisional
damping mechanisms such as Landau damping. A very successful fluid
model that approximates Landau damping in one dimension using
calculations in Fourier space was introduced by \citet{hammett-perkins:1990} and
\citet{hammett-dorland-perkins:1992}. It was later extended
(see e.g.~\citet{snyder-hammett-etal:1997} and \citet{passot-sulem:2003})
and is still relevant today for one-dimensional cases and
as a basis for other closures.

In two and three dimensions, it is a reasonable assumption that
damping mechanisms will drive temperature to a more isotropic
state. Fluid models that include a term which drives the pressure
tensor to isotropy were applied in collisionless reconnection
by \citet{hesse-winske-etal:1995} and \citet{johnson-rossmanith:2010}.
Later, a connection was made between the Landau damping fluid closures
and pressure isotropisation by \citet{wang-hakim-etal:2015}.
The result was the fluid model that was used for the mentioned
simulations of Ganymede's and Mercury's magnetospheres.
However, some problems became apparent; using this heat flux
closure in island coalescence reconnection, one does not
obtain the same reconnection rate scaling with system size
that is present in kinetic simulations \citep{ng-huang-hakim-etal:2015}.
It was also found that
the simple isotropisation closure does not support the LHDI when its
free parameter is chosen as in reconnection simulations.
Both issues can be addressed to some degree with a closure
expression that, like the original Hammett-Perkins approach,
incorporates calculations in Fourier space
\citep{ng-hakim-etal:2017,ng-hakim-etal:2019}.
Such a non-local closure, however, is not suitable
for large scale simulations because Fourier transforms and
the necessary global communication are
computationally extensive \citep{ng-hakim-etal:2019}.
Even better agreement with kinetic results can be achieved using
a local pressure or temperature gradient-driven closure,
as has been shown for both reconnection \citep{allmann-rahn-trost-grauer:2018}
and the LHDI \citep{ng-hakim-etal:2020}.

Besides Landau damping, particle trapping is another relevant
kinetic effect in collisionless reconnection. A fluid model that includes
particle trapping mechanisms was developed by \citet{le-egedal-etal:2009}
with an extension
of the classic equations of state. This closure yields good results
in guide field reconnection (also see \citet{egedal-le-daughton:2013,
le-daughton-etal:2016}) but is not designed for use in systems without a
guide field. In its current form, the model does not support
the LHDI \citep{le-stanier-etal:2019}.

In this paper, we simulate three-dimensional reconnection and current sheet
instabilities using the heat flux closure introduced in
\citet{allmann-rahn-trost-grauer:2018}. There, heat flux is assumed to be
proportional to the gradient of the pressure tensor's deviation from
isotropy. Thus, both the isotropisation character of the closely related closure
in \citet{wang-hakim-etal:2015} and the gradient dependence of the classic
Landau fluid closures are retained.

\section{Physical Models and Numerics}\label{sec:models}

A plasma is accurately described by distribution functions $f_s(\mathbf{x},\mathbf{v},t)$ for each species $s$.
The amount of particles located between $\mathbf{x}$ and $\mathbf{x} + \mathrm{d}\mathbf{x}$ with velocities
between $\mathbf{v}$ and $\mathbf{v} + \mathrm{d}\mathbf{v}$
is then given by $f(\mathbf{x},\mathbf{v},t)\,\mathrm{d}\mathbf{x}\,\mathrm{d}\mathbf{v}$. For collisionless
plasmas, the Vlasov equation determines the evolution of the distribution functions:
\begin{equation}
\frac{\partial f_{s}}{\partial t} +\mathbf{v} \cdot \nabla f_{s} + \frac{q_{s}}{m_{s}} (\mathbf{E + v \times B}) \cdot \nabla_{v} f_{s} = 0.
\label{eq:vlasov} \end{equation}
Various physical quantities can be obtained from the distribution function by taking moments. Multiplying by powers
of $\mathbf{v}$ and integrating over velocity space leads to expressions for particle density
$ n_s(\mathbf{x}, t) = \int f_s(\mathbf{x}, \mathbf{v}, t) \text{d}\mathbf{v}$, mean velocity
$ \mathbf{u}_s(\mathbf{x},t) = \frac{1}{n_s(\mathbf{x}, t)} \int \mathbf{v} f_s(\mathbf{x}, \mathbf{v}, t) \text{d}\mathbf{v}$,
pressure $\text{P}_{s} = m_{s} \int \mathbf{v}' \otimes \mathbf{v}' f_{s} \text{d}\mathbf{v}$
and heat flux $\text{Q}_{s} = m_{s} \int \mathbf{v}' \otimes \mathbf{v}' \otimes \mathbf{v}' f_{s} \text{d}\mathbf{v}$.
Here, $\otimes$ is the tensor product and $\mathbf{v}' = \mathbf{v - u}$.

The plasma quantities evolve according to the so called multi-fluid equations.
These can be derived by taking moments of the Vlasov equation and making a
physical assumption in order to close the resulting hierarchy. If moments up to
the pressure tensor are considered, the ten moment equations follow (named
like that because ten equations need to be solved):
\begin{equation}
\frac{\partial n_{s}}{\partial t} + \nabla \cdot (n_{s} \mathbf{u}_{s}) = 0 \, ,
\label{eq:tenmoment_continuity} \end{equation}

\begin{equation}
m_{s} \frac{\partial (n_{s} \mathbf{u}_{s}) }{\partial t} =
n_{s} q_{s} (\mathbf{E} + \mathbf{u}_{s} \times \mathbf{B}) - \nabla \cdot \mathcal{P}_{s} \, ,
\label{eq:tenmoment_movement} \end{equation}

\begin{equation}
    \frac{\partial \mathcal{P}^{s}_{ij}}{\partial t} - q_{s} (n_{s} u^{s}_{[i} E_{j]}
    + \frac{1}{m_{s}} \epsilon_{[ikl} \mathcal{P}^{s}_{kj]} B_{l})
    = - (\nabla \cdot \mathcal{Q}_{s})_{ij} \, .
\label{eq:tenmoment_energy} \end{equation}
$\mathcal{P}_{s} =  m_{s} \int \mathbf{v} \otimes \mathbf{v} f_{s} \text{d}\mathbf{v}$ and
$\mathcal{Q}_{s} =  m_{s} \int \mathbf{v} \otimes \mathbf{v} \otimes \mathbf{v} f_{s} \text{d}\mathbf{v}$ are
the second and third moment of the distribution function (multiplied by mass), $\epsilon_{ikl}$ is the
Levi-Civita symbol and the square brackets denote the sum over as many permutations of indices as
needed to make the tensors symmetric, for
example \[ u_{[i} E_{j]} = u_i E_j + u_j E_i. \]
This set of equations needs a closure approximation for the divergence of the heat flux which is
discussed in Sec.~\ref{sec:closure}. Heat flux $\mathrm{Q}$ is related to
the distribution function's moments according to
$\mathrm{Q}_{ijk} = \mathcal{Q}_{ijk} - u_{[i} \mathcal{P}_{jk]} + 2 m n u_i u_j u_k$.

The ten moment equations reduce to five moment equations, if the temperature
(and pressure, respectively) is isotropic.
Then Eq.~\eqref{eq:tenmoment_movement} becomes
\begin{equation}
m_{s} \frac{\partial (n_{s} \mathbf{u}_{s}) }{\partial t} =
n_{s} q_{s} (\mathbf{E} + \mathbf{u}_{s} \times \mathbf{B}) -
\frac{1}{3} \nabla (2 \mathcal{E}_{s} - m_s n_s u_s^2) -
\nabla \cdot (m_s n_s \mathbf{u}_s \otimes \mathbf{u}_s) \, ,
\label{eq:fivemoment_movement} \end{equation}
with scalar energy density $\mathcal{E}_s = \frac{m_{s}}{2} \int v^2 f_{s} \text{d}\mathbf{v}$.
Additionally assuming zero heat flux in this isotropic limit,
Eq.~\eqref{eq:tenmoment_energy} becomes

\begin{equation}
    \frac{\partial \mathcal{E}_{s}}{\partial t} +
    \frac{1}{3} \nabla \cdot (\mathbf{u}_s (5 \mathcal{E}_s - m_s n_s u_s^2)) -
    q_s n_s \mathbf{u}_s \cdot \mathbf{E} = 0.
\label{eq:fivemoment_energy} \end{equation}

Physical units in the simulations are normalised as follows:
Length is normalised over ion inertial length $d_{i,0}$ based on
density $n_0$, velocity over ion Alfv\'{e}n velocity $v_{A,0}$ based
on the magnetic field $B_0$, time over the inverse of the ion
cyclotron frequency $\Omega_{i,0}^{-1}$ and mass over ion mass $m_i$.
Further normalisations are vacuum permeability $\mu_0 = 1$
and Boltzmann constant $k_B = 1$.

The Vlasov simulations in this paper use the positive and
flux-conservative (PFC) method by \citet{filbet-sonnendruecker-etal:2001},
which is a semi-Lagrangian method, combined with backsubstitution
\citep{schmitz-grauer:2006b} for the velocity updates.
For the fluid equations we employ a centrally weighted essentially
non-oscillating (CWENO) method \citep{kurganov-levy:2000} and the
third-order Runge-Kutta scheme by \citet{shu-osher:1988}.
Maxwell's equations are solved by means of the finite-difference
time-domain (FDTD) method.

All of the simulations were performed
using the GPU accelerated \textit{muphy2} multiphysics simulation code
developed at the Institute for Theoretical Physics I,
Ruhr University Bochum. The code is designed for running kinetic,
fluid and hybrid schemes either individually or spatially coupled
to each other.

\section{Gradient-Driven Heat Flux Closure}\label{sec:closure}

A heat flux term that models kinetic dissipation
mechanisms needs to introduce damping into the ten moment
equations. This can be achieved, for example, with
an expression that takes the form of Fourier's law
\begin{equation}
\mathbf{q} = - n\,D\,\nabla T
\label{eq:fouriers_law}\end{equation}
with heat flux vector $\mathbf{q}$ and thermal conductivity $D$.
Heat flux approximations based
on temperature gradients have been used for plasmas
in the classic Braginskii equations already \citep{braginskii:1965}
and can also be the basis of Landau fluid closures.
\citet{hammett-perkins:1990} and \citet{hammett-dorland-perkins:1992}
used phase mixing theory to obtain in one-dimensional Fourier space
$\tilde{q} = - n_{0} \chi_{1} \sqrt{2} \frac{v_{t}}{|k|} i k \tilde{T}$
where tildes denote perturbed quantities and
the conductivity $D = \chi_{1} \sqrt{2} \frac{v_{t}}{|k|}$ is
dependent on the wave number $k$ and thermal velocity
$v_t$. This closure is suitable for
approximating kinetic Landau damping.

Fourier's law is trivially generalised to three dimensions
by replacing the heat flux vector with the third order heat
flux tensor $\mathrm{Q}$ and the scalar temperature with the
second order temperature tensor $\mathrm{T}$. Then the gradient
of the temperature tensor yields a third order tensor.
The conductivity from the Landau fluid closure can also easily
be transferred as it contains only the absolute value of the
wave number. To avoid calculations in Fourier space,
\citet{sharma-hammett-etal:2006} and \citet{wang-hakim-etal:2015}
introduced a simplification where
the continuous wave number field is replaced by a single typical wave
number $\mathbf{k}_0$, and we follow this approach.
Landau damping is the kinetic damping of plasma oscillations,
thus $1/|\mathbf{k}_{s,0}| = d_{s,0}$ is a natural choice,
$d_{s,0}$ being the species inertial length based on $n_0$.
We choose an initial ansatz $\mathbf{q} = - D\,\nabla P$
slightly different from Fourier's law which originates from
comparisons with kinetic simulations.
Since Landau damping is modelled, damping
is supposed to act on fast changing gradients due to wave-like
plasma activities and not on gradients from slowly changing
background quantities. Therefore, we subtract an equilibrium
pressure which is comparable to the consideration of perturbed quantities
like in the Hammett-Perkins closure.
For species $s$, the resulting heat flux then has the form
\begin{equation}
\mathrm{Q}_s = -\frac{\chi}{k_{s,0}}\,v_{t,s}\,\nabla\,(\mathrm{P}_s - \mathrm{P}_{s,0})
\label{eq:heat_flux_approximation_3d}\end{equation}
with the dimensionless parameter $\chi$ and an equilibrium pressure
$\mathrm{P}_{s,0}$. A Maxwell distribution is an equilibrium
distribution at which no Landau damping occurs and
the related pressure would then be isotropic. This motivation
is in accordance with studies where a closure term that drives
pressure towards isotropy led to good results 
\citep{hesse-winske-etal:1995,johnson-rossmanith:2010,wang-hakim-etal:2015}.
The closure produces wave damping, but due to the
approximation of wave number and pressure perturbation
it does not always give the correct Landau damping rate.
It is not obvious which kinetic effects are captured 
and to what degree,
for example \citet{hesse-winske-etal:1995} included an
isotropisation term to account for phase space instabilities.

We need an approximation of $\nabla \cdot \mathrm{Q}$ as a
closure to the ten moment equations
so we take the divergence. The derivative of $v_{t,s}$ is assumed to be
small and is neglected, and we obtain the Laplacian of the pressure tensor
which is simply the Laplacian applied to each of its components.
Then the final expression (hereafter called the gradient closure) is
\begin{equation}
\nabla \cdot \mathrm{Q}_{s} = -\frac{\chi}{k_{s,0}}\,v_{t,s}\,\nabla^{2}\,(\mathrm{P}_{s} - p_s\,\mathrm{id})
\label{eq:gradient_closure}\end{equation}
with $\chi = 3$ and $k_{s,0} = 1/d_{s,0}$. The isotropic pressure is given by
$p_s = (\mathrm{P}_{s,xx} + \mathrm{P}_{s,yy} + \mathrm{P}_{s,zz})/3$ and $\mathrm{id}$ denotes
the identity matrix. The thermal velocity is defined as $v_{t,s} = \sqrt{k_B T_s / m_s}$.
We find that $\chi = 3$ is a reasonable value in a broad range of plasma configurations,
but modifications to this dimensionless parameter can be made for further
improving the agreement with kinetic calculations in specialised
scenarios. Eq.~\eqref{eq:gradient_closure} reduces to the isotropisation closure
in \citet{wang-hakim-etal:2015} with the approximation $\nabla^2 \approx k_0^2$.
The gradient closure was first
used in \citet{allmann-rahn-trost-grauer:2018}.

For numerical stability and in order to enforce positive temperatures,
we set a floor for the diagonal elements of the
pressure tensor. The Laplacian is evaluated using second order central
finite differences. This explicit method may need smaller time
steps than the solution of the fluid equations themselves.
Therefore, the computation of the closure is subcycled with
a boundary exchange of the pressure tensor in between each subcycle.
The amount of subcycles necessary depends on the cell size and
the plasma configuration. In the three-dimensional reconnection
simulation in this paper, for example, eight subcycles were used.

\section{Two-Dimensional Reconnection}\label{sec:2d_reconnection}

\subsection{Harris Sheet}\label{sec:2d_reconnection_harris}

\begin{figure}
\includegraphics[width=\textwidth]{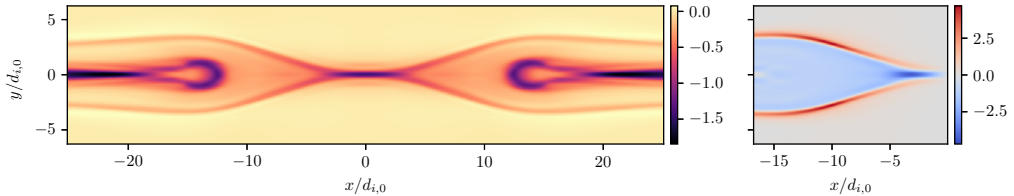}
\caption{Current and velocity profiles in two-dimensional
reconnection. Out-of-plane current $j_{z} / (n_0 v_{A,0})$
(left) and electron outflow velocity $u_{x,e} / v_{A,0}$ (right) at
$t = 65\,\mathrm{\Omega_{i,0}^{-1}}$.}
\label{fig:rec_2d_65}\end{figure}

Magnetic reconnection can develop from a perturbation of the Harris sheet
\citep{harris:1962} equilibrium. As initial conditions for the simulation
we use parameters similar to the GEM (Geospace Environmental Modeling)
reconnection setup \citep{birn-drake-shay-etal:2001} but with
a larger domain of size
$L_{x} \times L_{y} = (32\pi\times8\pi)\,d_{i,0} \approx (100 \times 25)\,d_{i,0}$,
resolved by $2400 \times 600$ cells.
Further differences are in the ion to electron mass ratio which is set to
$m_i / m_e = 100$ and in the speed of light $c = 30\,v_{A,0}$.
The domain is periodic in $x$-direction, has conducting walls for fields
and reflecting walls for particles in $y$-direction and is translationally
symmetric in $z$-direction. The initial configuration of the magnetic field is
$B_{x}(y) = B_{0} \tanh(y/\lambda)$ and the particle density is
$n(y) = n_{0} \sech^{2}(y/\lambda) + n_{b}$ with $\lambda = 0.5\,d_{i,0}$ and
background density $n_{b} = 0.2\,n_0$. Temperature is defined by
$n_{0} k_{B} (T_{e}+T_{i}) = B_{0}^{2} / (2 \mu_{0})$, $T_{i}/T_{e} = 5$.
A perturbation of the magnetic field is added to break the equilibrium which
takes the form $\mathbf{B} = \hat{\mathbf{z}} \times \nabla \psi$ where the
perturbation in the magnetic flux is given by
$\psi(x,y) = 0.1 \cos(2 \pi x / L_{x}) \cos(\pi y / L_{y}) B_0 d_{i,0}$.
In the Harris equilibrium, the current resulting from the magnetic field
configuration is distributed among electrons and ions according
to $u_{z,i}/u_{z,e} = T_i/T_e$.

The ten moment gradient simulation shown in Fig.~\ref{fig:rec_2d_65} has
an x-point profile comparable to existing particle-in-cell simulations
with large system sizes.
The electron outflow jet near the x-point is visible in the right plot in dark
blue. Its structure matches the kinetic results shown in Fig.~2 of
\citet{nakamura-nakamura-etal:2018}, including the in- and outflow regions
along the separatrix boundary. \citet{nakamura-genestreti-etal:2018} modelled
a reconnection event detected by MMS where the outflow jet is similar.
A difference between kinetic and ten moment fluid models is that
the onset of reconnection can take significantly longer in fluid simulations
depending on the heat flux closure
\citep{wang-hakim-etal:2015,allmann-rahn-trost-grauer:2018}. This is
also the case in the present simulation, and although onset duration
is not a particularly important aspect in two-dimensional Harris sheet
reconnection (as current sheet formation is ignored anyway), it is
important for the interaction with current sheet
instabilities in three-dimensional simulations.

After reconnection, a secondary island evolves at the x-point location. If
perturbations are applied to break symmetry, this island is ejected, but
in the perfectly symmetric case shown here, the island stays in the middle
of the domain and two new x-points are created. When the current sheet
enlarges in x-direction, at later times secondary islands also
evolve in kinetic simulations with zero guide field
\citep{karimabadi-daughton-etal:2007,klimas-hesse-etal:2008},
but more often when a guide field is present \citep{drake-swisdak-etal:2006}.
Ion-scale islands, also referred to as flux transfer events, have been
observed in magnetosphere \citep{hasegawa-kitamura-etal:2016, hwang-sibeck-etal:2016}.
Nevertheless, the ten moment gradient model seems to be more susceptible
to formation of secondary islands.

\subsection{Island Coalescence}\label{sec:2d_reconnection_isl}

\begin{figure}
\centering\includegraphics[width=0.8\textwidth]{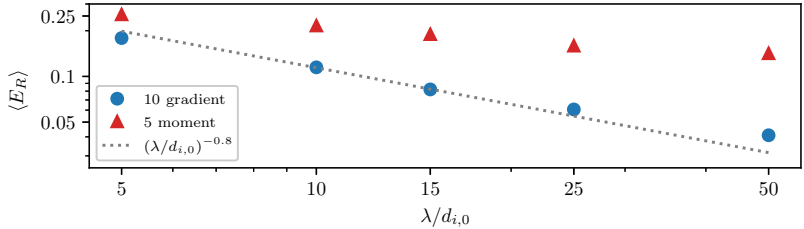}
\caption{Scaling of average reconnection
rate with system size in island coalescence reconnection:
ten moment gradient and five moment simulations.}
\label{fig:isl_reconnection_rates}\end{figure}

Reconnection studies are often initialised with a perturbed current sheet
as described in the previous section. This approach results in reconnection rates and
an x-point structure comparable
to reconnection events in the magnetosphere. The procedure of reconnection onset,
however, differs from that in nature since the formation of the
current sheet is not considered. A model that can be applied to
the magnetosphere should also be able to capture the physics of current sheet
formation appropriately. This can be tested using the island coalescence reconnection
setup where kinetic simulations showed lower reconnection rates
for larger system sizes \citep{stanier-daughton-etal:2015}, and the ten moment
fluid model with the gradient closure can reproduce the reconnection rate scaling
\citep{allmann-rahn-trost-grauer:2018}. The correct time development of reconnection
is particularly important for the interplay with current sheet instabilities like
the LHDI. In order to see if the kinetic results can be reproduced using the parameters chosen here
for Harris sheet reconnection and the lower-hybrid drift instability,
we perform island coalescence simulations for different system sizes with
$\chi = 3$ in Eq.~\eqref{eq:gradient_closure}.

The initial geometry is defined by density
$n = n_{0} (1 - \epsilon^{2}) / (\cosh(y/\lambda) + \epsilon \cos(x/\lambda))^{2} + n_{b}$
and magnetic potential
$A_{z} = - \lambda B_{0} \ln( \cosh(y/\lambda) + \epsilon \cos(x/\lambda)$ with
a system size of $L_{x} \times L_{y} = (2 \pi \lambda \times 4 \pi \lambda)\ d_{i,0}$.
For a more detailed description of the setup see \citet{allmann-rahn-trost-grauer:2018}
(also \citet{stanier-daughton-etal:2015}, \citet{ng-huang-hakim-etal:2015}).
Reconnection rate is evaluated using the magnetic flux, which is the
integral over $B_x$ from the O- to the X-point $\Psi = \int^{\mathrm{X}}_{\mathrm{O}}\,B_{x}\,\mathrm{dy}$
and is normalised over $B'$, the maximum of the magnetic field's absolute value
at $x=0$ and time $t = 0$. With $v'_{A} = B'/\sqrt{\mu_{0} n_{0} m_{i}}$, the
normalised reconnection rate is $E_{R} = \frac{\partial \Psi}{\partial t} / (B' v'_{A})$.

In Fig.~\ref{fig:isl_reconnection_rates} the average reconnection rate is shown depending
on the parameter $\lambda$ which determines the system size. The dotted line
represents the scaling of $\langle E_R \rangle$ with $\lambda$
found by \citet{stanier-daughton-etal:2015}
in kinetic particle-in-cell simulations up to $\lambda = 25\,d_{i,0}$, and
the ten moment gradient model comes close to
$\langle E_R \rangle \propto (\lambda / d_{i,0})^{-0.8}$. Slightly better agreement
can be achieved by tuning $\chi$ and $k_{s,0}$, but it is important that the closure works
fine with the same parameters in different configurations -- and that seems
to be the case for $\chi = 3$ and $k_{s,0} = 1/d_{s,0}$.
In addition to the ten moment gradient results, we show
performance of the five moment model, which is able to capture the lower-hybrid
drift instability (cf.\ Sec.~\ref{sec:lhdi}) and was not considered in
previous studies of island coalescence. The drop of reconnection rate is much smaller
and the scaling with system size is only somewhat stronger than that of Hall-MHD in
\citet{stanier-daughton-etal:2015}. The ten moment
model using the isotropisation closure or a non-local closure outperform the
five moment model, although they do not reach the scaling of the kinetic and
ten moment gradient models \citep{ng-hakim-etal:2017,allmann-rahn-trost-grauer:2018}.

\section{Lower-Hybrid Drift Instability (LHDI)}\label{sec:lhdi}

\begin{figure}
\centering \includegraphics[width=0.9\textwidth]{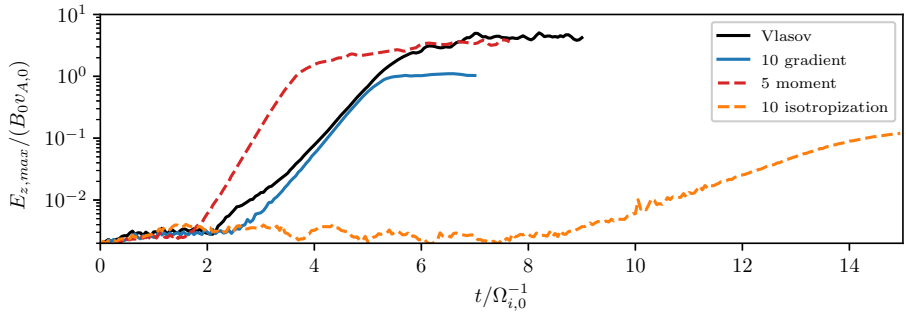}
\caption{Time development of the supremum norm of $E_{z}$ in LHDI simulations ($m_i/m_e = 100$).}
\label{fig:lhdi_e_max_time}\end{figure}

When magnetic reconnection is simulated in two dimensions, the assumed
approximation is that the plasma is homogeneous along the out-of-plane
dimension. Thin current sheets, as they are present in reconnection, are often
susceptible to instabilities though and these can cause highly
turbulent features in the out-of-plane dimension. The initial setup for a
two-dimensional analysis of various current sheet instabilities can again
be a Harris sheet equilibrium. The connection to the two-dimensional
reconnection setup is that instead of the $x$-$y$-plane,
now the $y$-$z$-plane is simulated for one fixed point in $x$-direction.
So the initial configuration is again $n(y) = n_{0} \sech^{2}(y/\lambda) + n_{b}$,
$B_{x}(y) = B_{0} \tanh(y/\lambda)$,
$n_{0} k_{B} (T_{e}+T_{i}) = B_{0}^{2} / (2 \mu_{0})$ and
$c = 30\,v_{A,0}$. For the velocity profiles it follows that
$u_{z,s} = 2 T_{s,0} B_0 \sech^{2}(y/\lambda) / (q_s n_s \lambda \mu_0)$.
The instabilities are initiated by adding perturbations which can either be random noise
or explicit modes that are supposed to be examined. In PIC simulations,
the particle noise also excites instabilities and when single precision
calculations are used (especially combined with the FDTD method) even
the noise from machine precision drives instabilities.
Therefore we utilise double precision calculations in all cases
in order to avoid unphysical results.
There are four important aspects to consider when studying the LHDI:
the growth rate, the duration of onset, the saturation
and the potential excitation of secondary instabilities. These characteristics
are most influenced by current sheet thickness, background density and
ion-electron temperature ratio. Generally, the growth rate is higher for
thin current sheets, low background density and high $T_i/T_e$.

\subsection{Electrostatic Branch}\label{sec:lhdi_electrostatic}

The electrostatic LHDI is often the
first current sheet instability to arise and generates fluctuations
located at the edges of the current sheet.
For studying the electrostatic LHDI branch we choose parameters as in
\citet{ng-hakim-etal:2019} so that there is the possibility of
comparing the results. The parameters are: half-thickness of the
current sheet $\lambda = \rho_{i,0} \approx 0.67 d_{i,0} $, background density
$n_b = 0.001\,n_0$ and temperature ratio $T_{i,0}/T_{e,0} = 10$,
where $\rho_{i,0} = \sqrt{k_B T_{i,0} / m_i} / \Omega_{i,0}$ is the
ion gyroradius. These are suitable conditions for a fast
development of the LHDI. \citet{ng-hakim-etal:2019} used a
mass ratio of $m_i/m_e = 36$ which is interesting for
magnetosphere modelling where reduced mass ratios are common.
Nevertheless, it is necessary to consider more realistic
mass ratios, both for obtaining physically relevant results and
for comparing numerical models.
In order to analyse the influence of mass ratio,
we perform simulations using different electron masses.
Perturbations that exist in realistic plasmas can be
modelled by adding random noise to the initial setup.
That way, the contribution and interaction of all modes
as well as the occurence of other instabilities than the LHDI
can be observed. For the simulations in this subsection,
linearly distributed random noise is added to the initial $B_{x}$
with a magnitude of $10^{-4} B_0$.
The simulated domain is of size
$L_y \times L_z = (12.8 \times 6.4) \rho_{i,0} \approx (8.6 \times 4.3) d_{i,0}$
with a spatial resolution of $512 \times 256$ for $m_i/m_e \leq 100$
and $1024 \times 512$ for higher mass ratios. In the Vlasov
simulations, the extent of velocity space is $\pm 8 v_{t,s}$ resolved
by $2.1$ cells per $v_{t,e}$. In the $m_i/m_e = 250$ case, for example,
electron velocity space goes from $-27 v_{A,0}$ to $27 v_{A,0}$ and
ion velocity space from $-5 v_{A,0}$ to $5 v_{A,0}$ with
a total resolution of $1024\times512\times114^2$ cells.

\begin{wrapfigure}{R}{0.45\textwidth}
\includegraphics[width=0.45\textwidth]{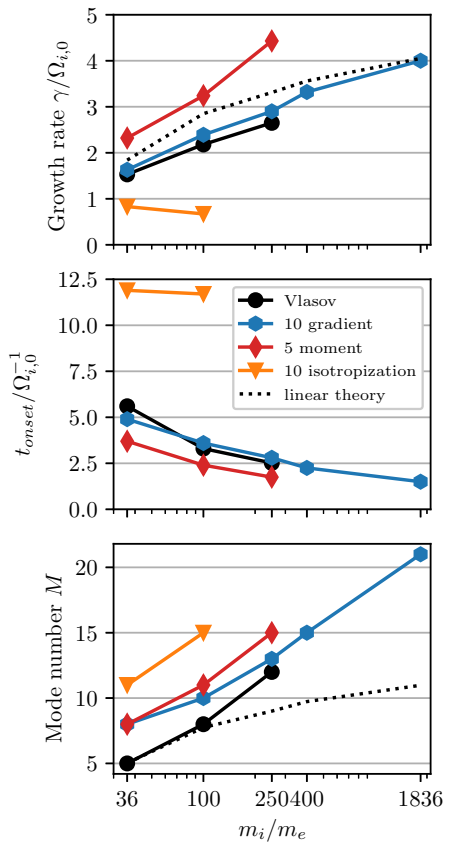}
\caption{Scaling of LHDI characteristics with mass ratio for different
models. Growth rate of the instability (upper), onset duration (middle)
and number of modes (lower).}
\label{fig:lhdi_mass_scaling}\end{wrapfigure}

Growth rate, onset duration and the mode number that develops
are shown in Fig.~\ref{fig:lhdi_mass_scaling} for the Vlasov
model, the ten moment model using the gradient closure, the
five moment model and the ten moment model using the isotropisation
closure in \citet{wang-hakim-etal:2015}. Growth rate $\gamma$
is determined via a fit in the phase where
$||E_z||_{\infty} \propto \exp(\gamma t)$ and we define the
instability's onset duration as the time until
$||E_z||_{\infty} \geq 0.02\,B_0 v_{A,0}$ which was the value
where exponential growth started in the simulations.

The growth rate agrees reasonably well between the kinetic
Vlasov model and the ten moment gradient model. For the
$m_i/m_e = 100$ case, time development of the electric field
is shown in Fig.~\ref{fig:lhdi_e_max_time} where the curves'
slopes correspond to the growth rates. The growth rates
are $2.2\,\Omega_{i,0}$ and $2.4\,\Omega_{i,0}$ for the Vlasov
and ten moment gradient models, respectively, whereas growth
rate is higher in the five moment model ($3.2\,\Omega_{i,0}$)
and significantly lower in the ten moment isotropisation model
($0.67\,\Omega_{i,0}$). The latter approximates the divergence
of heat flux as
$\nabla \cdot \mathrm{Q}_{s} = k_{s,0}\,v_{t,s}\,(\mathrm{P}_{s} - p_s\,\mathrm{id})$
and we choose $k_{s,0} = 1/d_{s,0}$ for the free parameter which was used
in reconnection studies. When $k_{s,0} \rightarrow \infty$,
the isotropisation closure approaches the five moment limit (isotropy)
and the growth rate increases accordingly. In this case, however,
reconnection cannot be modelled well. Vlasov simulations with
high mass ratios are not feasible, but we performed ten moment
gradient simulations with realistic mass ratio, and the growth
rate goes up to $4\,\Omega_{i,0}$. This should be considered
when interpreting magnetosphere simulations where
severely reduced mass ratios such as $m_i/m_e = 25$ are 
employed.

\begin{figure}
\includegraphics[width=\textwidth]{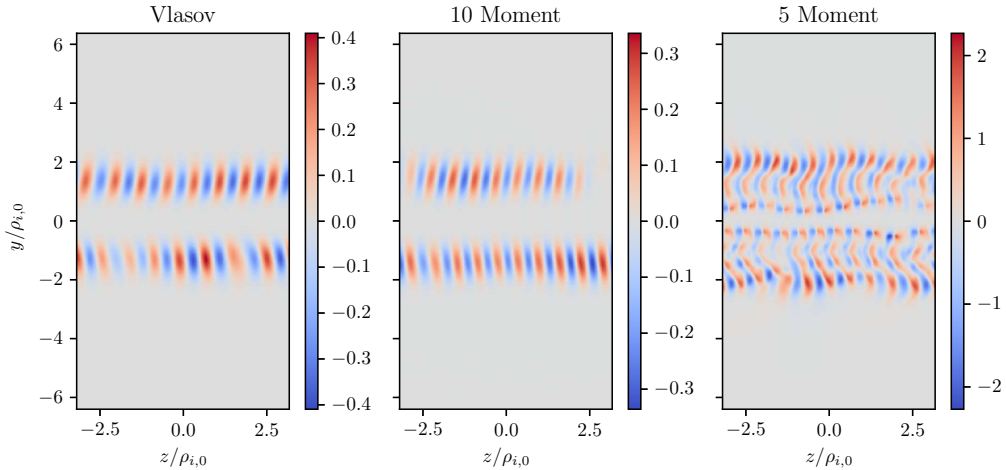}
\caption{$E_z/(B_0 v_{A,0})$ in simulations of the LHDI at $t = 4.75\,\Omega_{i,0}^{-1}$ for
the Vlasov model, the ten moment gradient model and the five moment model ($m_i/m_e = 100$).}
\label{fig:lhdi_color}\end{figure}

The onset duration is, for Harris sheet reconnection
simulations, as relevant as the growth rate since the
influence of the LHDI at a certain time results from
the combination of both. It is important to note
that the onset duration depends heavily
on the magnitude of the initial perturbation. But
when the perturbation is fixed,
the duration until exponential growth starts can
be examined for the various models.
In Fig.~\ref{fig:lhdi_color}
the structure and magnitude of the electric field at
a certain time is shown, which gives an impression of
the combined effect of onset duration and growth rate. The
difference between the ten moment gradient model and
the Vlasov model is small and the slightly higher
E-field value in the Vlasov case results from the earlier
onset. The five moment model is already in a state
where the fastest growing mode has saturated and
secondary modes are excited. This rapid development can be
attributed more to the very fast onset than to
the higher growth rate. The five moment results are
in contrast to the findings of \citet{ng-hakim-etal:2019}
who reported a much too slow LHDI growth using
this model. They agree, however, with a
study by \citet{tenbarge-ng-etal:2019} who modelled
an MMS reconnection event and found the five moment
model to be not well-suited due to explosive
growth of the LHDI. Our findings are also consistent
with the $k_{s,0} \rightarrow \infty$ limit of the
ten moment isotropisation closure.
The onset time in the ten moment isotropisation model using
normal parameters is very late,
decreases with mass ratio, and for $m_i/m_e=250$
no instability grows at all during the simulated time
span.

The mode number shown in Fig.~\ref{fig:lhdi_mass_scaling} is
defined as the $M=k_z L_z/(2 \pi)$ that develops out of
random noise in the respective simulation and thus corresponds
to the fastest growing mode (or one of the fastest, if
growth rates are very close for several modes). The mode
numbers do not agree between the Vlasov and fluid models
for low mass ratios. While the number of modes itself is of
little relevance for reconnection, it is crucial to always look at the
respectively fastest growing modes when comparing different models.
\citet{ng-hakim-etal:2019} explicitly initialised
the $M=8$ mode in the $m_i/m_e=36$ case which is the fastest mode for
the fluid models but a
rather slow mode in the Vlasov model (we
obtain a growth rate of $1.1\,\Omega_{i,0}$ for the
$M=8$ mode in agreement with \citet{ng-hakim-etal:2019}).
Therefore, only looking at a single mode
is not an appropriate method of measuring the capability of
fluid models to reproduce the kinetic LHDI.
At mass ratio $m_i/m_e=250$ the Vlasov
and the ten moment gradient model results become closer
and mode numbers keep increasing towards realistic mass
ratio in the fluid simulations.

It is interesting to compare the simulation results
to linear kinetic theory, especially at high mass ratios where the
Vlasov simulations are not available for validating the fluid
model. An expression for the local dispersion relation of the
LHDI according to linear kinetic theory is given in
Eq.~(52) of \citet{davidson-gladd-etal:1977}. The dispersion
relation can be solved numerically to give the dependence of growth
rate on the wave number. The local density and magnetic
field values were taken at the $y$-location of the fastest modes
in the simulations. The results from linear theory are 
shown in Fig.~\ref{fig:lhdi_mass_scaling} with dotted lines. The growth rate agrees well
between theory and the non-linear Vlasov simulation at $m_i/m_e = 36$
but the theory estimates slightly higher growth rates than the
kinetic simulation at more realistic mass ratios.
This suggests that non-linear electron effects play an important role.
In the theoretic prediction of the mode number the
discrepancy is more evident; there is excellent agreement
between kinetic theory and the Vlasov simulations at low mass ratios,
but at $m_i/m_e = 250$ the
theory does not reproduce the higher mode number from the
simulation. When the Vlasov simulation results
are not available, the gradient fluid model closely matches the growth
rates predicted by linear kinetic theory. Concerning the
mode number of the fastest growing mode, the theory yields
better results at low mass ratios, but the fluid simulation can
capture non-linear effects and the interaction between various modes leading to more
reliable results when realistic mass ratio is approached.

The mass ratio dependence of LHDI onset and growth rate the simulations supports our choice of
$k_{s,0} = 1/d_{s,0}$ in the gradient closure. More heat flux due
to the closure leads to increased LHDI growth rates and earlier onset.
Since $d_{s,0} \propto \sqrt{m_s}$, the higher $k_{s,0}$ compensates the
increased heat flux caused by $v_{t,s}$ in the closure at small
electron masses and prevents the electrons
from being thermalised. If
$k_{s,0}$ is chosen to be independent of mass, the LHDI grows too
fast at high mass ratios, whereas with $k_{s,0} = 1/d_{s,0}$
there is good agreement between kinetic and ten moment gradient
results.

\subsection{Electromagnetic Branch}\label{sec:lhdi_electromagnetic}

\begin{figure}
\centering \includegraphics[width=0.9\textwidth]{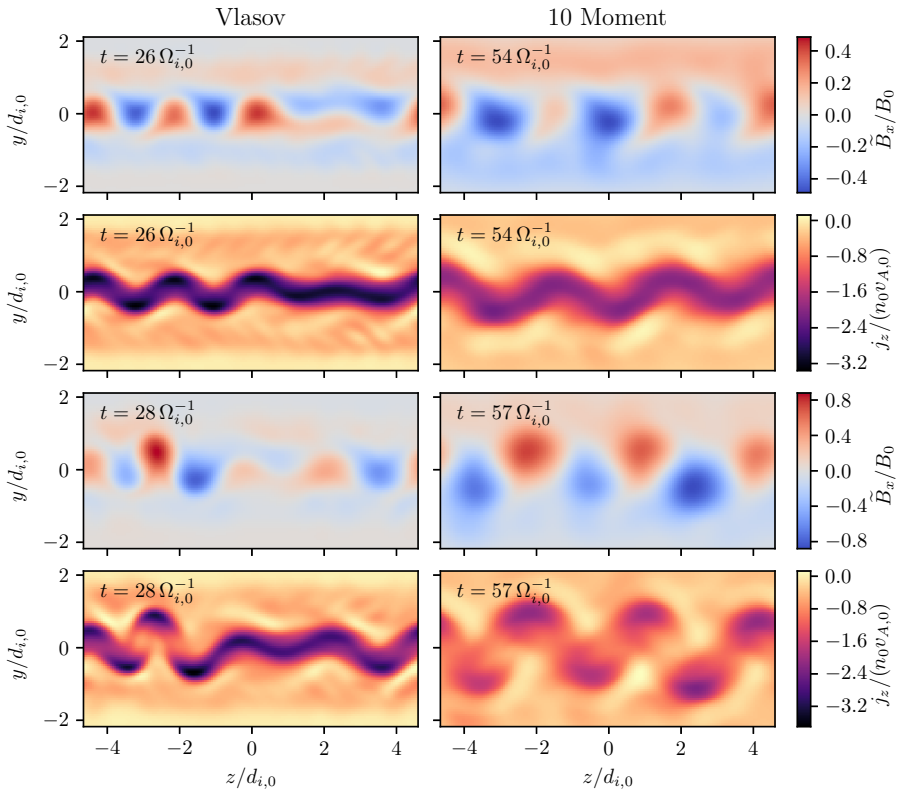}
\caption{Electromagnetic mode of the lower-hybrid drift instability in
Vlasov and ten moment gradient simulations. Shown are the deviation
of the magnetic field from the initialized profile
$\widetilde{B}_x = B_x - \tanh(y/\lambda)$ and the out-of-plane
current density $j_z$.}
\label{fig:lhdi_em_gem}\end{figure}

The electromagnetic LHDI has lower growth rates compared to the
electrostatic branch, but it penetrates into the center of the
current sheet and thus can potentially have more significant
impact on (for example) reconnection. Therefore, we test whether
the gradient fluid model can also reproduce these
more complex electromagnetic modes. A detailed analysis of the
electromagnetic branch using kinetic theory and simulations
can be found in \citet{daughton:2003}. Since we are interested
in the effect of the electromagnetic LHDI within reconnection,
we choose parameters and resolutions as in the fluid simulation
of three-dimensional reconnection that is discussed in the next section.
The parameters are $n_b = 0.2\,n_0$, $T_i/T_e = 5$, $\lambda = 0.5\,d_{i,0}, m_i/m_e=100$
and initial random noise added to electron density with
a magnitude of $\delta n_e = 10^{-8}\,n_0$. It should be noted
that electromagnetic LHDI branch is also present in Vlasov and fluid
simulations when the setup from the previous section is used. The domain is now of
size $L_y \times L_z = (6 \pi \times 2 \pi) d_{i,0}$ resolved by
$450\times150$ cells.

In Fig.~\ref{fig:lhdi_em_gem} the fluctuations in the magnetic field
caused by the electromagnetic LHDI are shown for kinetic Vlasov and ten moment
fluid simulations next to the current sheet density to demonstrate the influence of the mode.
Onset duration of both the elctrostatic and the
electromagnetic LHDI modes is longer using the fluid model compared
to the kinetic simulation with these plasma parameters. Although
an accurate representation of onset times in all parameter regimes
would of course be desirable, this behaviour is consistent with
the longer reconnection onset of the fluid model that was mentioned
in Sec.~\ref{sec:2d_reconnection}. Growth rate of the electromagnetic
modes that are shown in Fig.~\ref{fig:lhdi_em_gem} (concerning the amplitude of magnetic field fluctuations)
is $\gamma \approx 0.41\,\Omega_{i,0}$ in the Vlasov simulation and
$\gamma \approx 0.21\,\Omega_{i,0}$ in the fluid simulation.
In order to still compare the influence of the mode in both models,
we choose times in the plot when the magnetic fluctuations
have a similar magnitude.
The kinetic Vlasov solution has the typical spatial
structure of the electromagnetic mode at $t = 26\,\Omega_{i,0}^{-1}$
which is not perfectly reproduced by the ten moment gradient model
($t = 54\,\Omega_{i,0}^{-1}$),
but in the latter the mode does also arise in the center of the current sheet
and the wave length is similar. At this stage the electromagnetic
LHDI already causes significant kinking of the current sheet and
the magnitude of the magnetic field fluctuations is high. In
contrast to the findings in \citet{daughton:2003} the saturation
is not at a moderate level. Instead, the fluctuations in the magnetic field keep growing
so that $\widetilde{B}_x > 1\,B_0$, causing significant modifications to the current sheet and
magnetic field profile as can be seen in the lower panels of Fig.~\ref{fig:lhdi_em_gem}
(times $t = 28\,\Omega_{i,0}^{-1}$ and $t = 57\,\Omega_{i,0}^{-1}$ respectively).
The reason for the high saturation level is unclear, but of course the
results cannot be directly compared to \citet{daughton:2003} because
the plasma parameters are different.
In Harris sheet reconnection simulations, the reconnection process is
started by an initial perturbation to the magnetic field. If
strong fluctuations in the magnetic field due to the electromagnetic LHDI arise
before reconnection, they can also initiate reconnection leading to
multiple x-lines and turbulence.
The high saturation level of the LHDI is surprising, considering that many studies
suggest that very strong kinking of magnetosphere current sheets
is primarily caused by slower instabilities such as the
ion-ion kink instability or Kelvin-Helmholtz type instabilities
\citep{lapenta-brackbill:2002,lapenta-brackbill-etal:2003,baumjohann-roux-etal:2007}.
A detailed analysis of the saturation of the electromagnetic LHDI branch
and its effect on current sheet stability using kinetic simulations
in various parameter regimes goes beyond the scope of this paper
but is a subject of future research.

The fluid model cannot excactly represent the complex electromagnetic
LHDI modes, but the key properties and the effect on the current sheet
are in agreement with the Vlasov simulation including high
saturation levels of the magnetic field fluctuations.

\section{Three-Dimensional Reconnection}\label{sec:3d_reconnection}


\begin{figure}
\centering
\includegraphics[width=\textwidth]{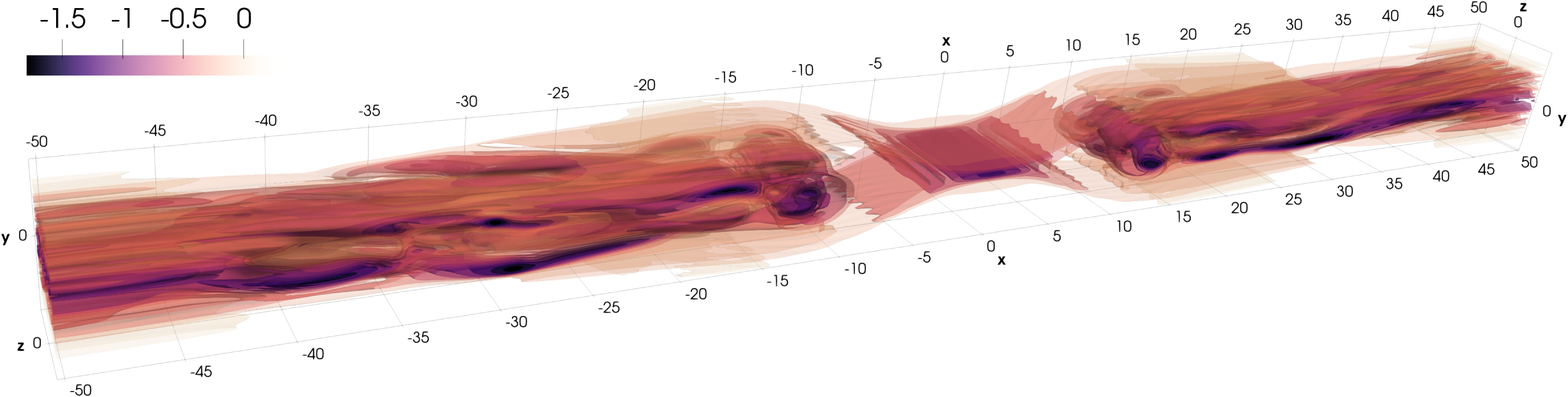}\\
\vspace{0.5cm}
\includegraphics[width=\textwidth]{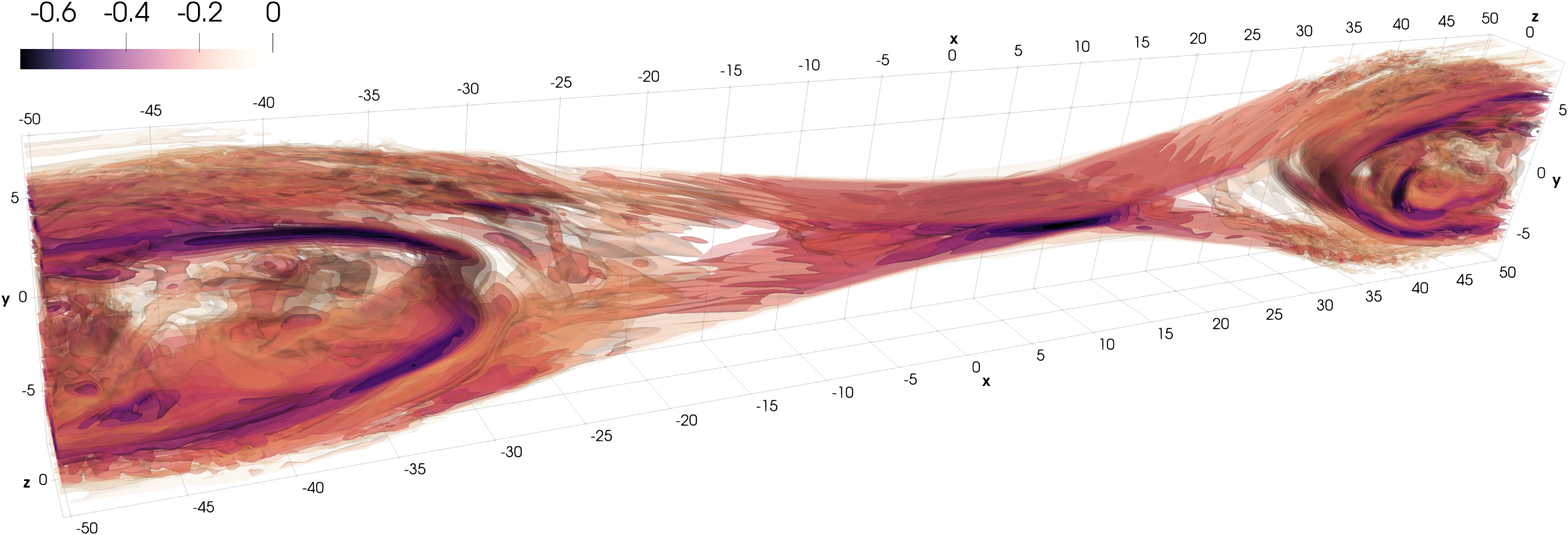}
\caption{Contour plots of $j_{z} / (n_0 v_{A,0})$ 
from the three-dimensional reconnection simulation
at times $t = 65\,\Omega_{i,0}^{-1}$ (upper)
and $t = 95\,\Omega_{i,0}^{-1}$ (lower).}
\label{fig:3d_rec_color}\end{figure}

In order to examine the effect of the LHDI on reconnection using the fluid model,
we study the interplay of the two in a three-dimensional setup that is identical
to the one of the Harris sheet reconnection simulation in
Sec.~\ref{sec:2d_reconnection_harris} -- except for the additional dimension
and a tiny initial perturbation of electron density. The domain now has the size 
$L_{x} \times L_{y} \times L_{z} = (32\pi \times 8\pi \times 2\pi)\,d_{i,0}
\approx (100 \times 25 \times 6)\,d_{i,0}$ with a resolution of
$2400 \times 600 \times 150$ cells and periodic boundary conditions
in $z$-direction. Linearly distributed random noise is added to the
initial electron density with a maximum magnitude of $\delta n_e = 10^{-8} n_0$
so that instabilities like the LHDI can develop. The perturbation's
magnitude influences the LHDI's onset duration and therefore has great impact
on the effect of the LHDI, as will be discussed later. Using a small random
perturbation of electron density, we give current sheet instabilities
maximum freedom in their development.

Contour renderings of the current density $j_{z}$ are given in Fig.~\ref{fig:3d_rec_color}
after the reconnection current sheet has formed and at a later stage when
turbulence dominates. The upper plot shows the same time as Fig.~\ref{fig:rec_2d_65}
for the two-dimensional case and similar features are visible with a slightly
smaller x-line current sheet in the three-dimensional simulation.
Fluctuations in the third dimension are moderate
in the electron diffusion region but strong in the outflow where
the electromagnetic LHDI
has impaired the current sheet. The electrostatic LHDI mode also affects the current density
and can be seen in the contours at $t=65\,\Omega_{i,0}^{-1}$ left and right
of the x-line. Around $t=68\,\Omega_{i,0}^{-1}$ a
secondary island (plasmoid) forms and is ejected in the negative x-direction.
Later, the electron diffusion region becomes more turbulent and the x-line
current sheet broadens, as is visible in the lower panel.

\begin{wrapfigure}{l}{0.5\textwidth}
\includegraphics[width=0.5\textwidth]{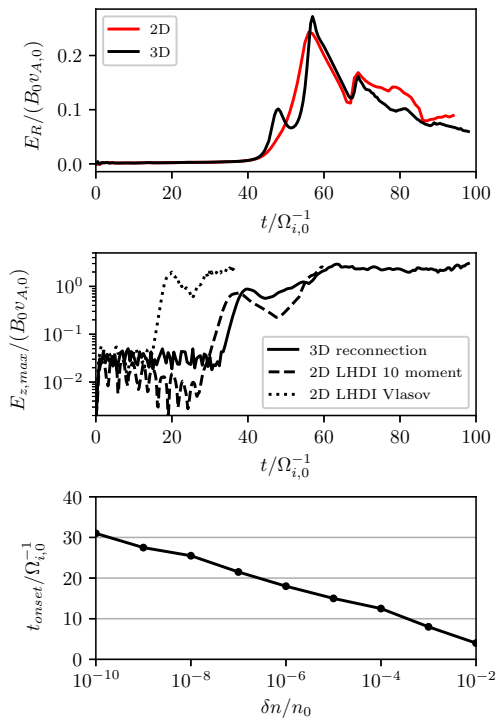}
\caption{Reconnection rate (upper),
supremum norm of $E_z$ in 
three-dimensional reconnection compared
to two-dimensional LHDI (middle)
and LHDI onset duration depending
on initial perturbation (lower).}
\label{fig:rec_rate_lhdi_development}
\end{wrapfigure}

\begin{figure}
\centering\includegraphics[width=\textwidth]{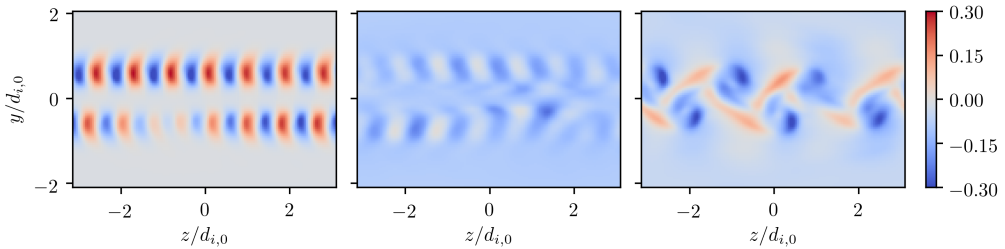}
\caption{Instabilities at the x-line in 3D reconnection.
Shown are y-z-slices of $E_z/(B_0 v_{A,0})$
at times $t=42\,\Omega_{i,0}^{-1}$ (left),
$t=54\,\Omega_{i,0}^{-1}$ (middle) and
$t=98.5\,\Omega_{i,0}^{-1}$ (right).}
\label{fig:lhdi_cssi_3d}\end{figure}

The reconnection rate is shown in Fig.~\ref{fig:rec_rate_lhdi_development} (top) for
both 2D and 3D, evaluated by averaging the value of $E_z$ at the x-line
over $z$.
We define the x-line as the location in the $x$-$y$ plane
where $B_y$ averaged over $z$ goes through zero to the left (i.e.\ negative $x$-direction)
of the maximum of $B_y$. When two x-lines are present due to a
magnetic island, this corresponds to the one at larger $x$-position as
the island is ejected in negative $x$-direction in the three-dimensional
simulation. The island formation is visible in the form of a bump around
$t=68\,\Omega_{i,0}^{-1}$ and is associated with an increase
of reconnection rate. The island is not
ejected in the two-dimensional simulation with perfect symmetry
which may cause the slightly higher reconnection rate at later times
in the 2D run.
The earlier start of fast reconnection in
three dimensions can likely be attributed to current sheet
thinning and electron heating caused by the LHDI. After fast
reconnection has started, the increased electron temperature
and weaker density gradient
in the electron diffusion region leads to a decay of the LHDI
which may be the reason for the temporary drop of the
reconnection rate at $t=48\,\Omega_{i,0}^{-1}$. The development
of instabilities at the x-line is demonstrated in
Fig.~\ref{fig:lhdi_cssi_3d} at three exemplary points in time.
The LHDI (left) is weaker and arises later at the x-line than 
in the rest of the current sheet and when fast reconnection has
started it decays and leaves moderate turbulence behind (middle)
caused by secondary modes. In the later stage, a different
instability emerges (right) which has common features with
the current sheet shear instability (CSSI) shown in
Fig.~3 of \citet{fujimoto-sydora:2017} concerning wave
length, $y$-extent and magnitude. To confirm whether the instability in
our simulation is indeed the CSSI will be left for
future work as it is not the primary subject of this study.
However, the structure of the electric field at the x-line does
agree with the kinetic PIC simulation of three-dimensional
reconnection in \citet{fujimoto-sydora:2017} although
there, the turbulence is much stronger around the x-line.
In contrast, \citet{nakamura-genestreti-etal:2018} found negligible
three-dimensionality in the electron diffusion region. This
is not necessarily a contradiction and is likely caused
by the different background densities which were $0.044 n_0$
in the former and $0.3 n_0$ in the latter study attended
by respectively fast or slow growth of the LHDI. The
low temperature ratio of $T_i/T_e = 3$ in \citet{nakamura-genestreti-etal:2018}
may have further slowed down the LHDI development. Our setup with $n_b=0.2 n_0$
and $T_i/T_e = 5$ lies between the two which is also
evident in the amount of turbulence present.
Recent studies on the role turbulence
and the LHDI in reconnection have discussed
a possible increase in reconnection rate due to modifications
in the Ohm's law caused by turbulence around
the x-line (so called anomalous dissipation). The
focus was on the electromagnetic LHDI in the reconnecting
current sheet and there are findings that support the
importance of turbulence \citep{price-swisdak-etal:2016,price-swisdak-etal:2017}
and some that estimate a rather low influence of
anomalous dissipation \citep{le-daughton-etal:2017,le-daughton-etal:2018}.
The small electromagnetic fluctuations at the x-line
in our three-dimensional fluid simulation at earlier
time (Fig.~\ref{fig:lhdi_cssi_3d} middle) do have a
similar structure as those shown in Fig.~7b of \citet{price-swisdak-etal:2017}
but in our case contributions of the electromagnetic
LHDI at the x-line decay fast because of the low density
gradient. Therefore, the LHDI can not make long-term contributions
apart from inducing the later CSSI-like instability. It should be
noted that this could potentially be different in the reconnection
scenario that aforementioned studies used where density and
temperature asymmetries are present. The possible influence
of the CSSI-like instability on reconnection rate will be
an interesting topic of future research.

For agreement of the fluid simulations of
three-dimensional reconnection with published kinetic PIC simulations,
the relation between instability onset duration
and reconnection onset duration is a central issue since both
depend on the closure approximation. If reconnection onset takes too
long with respect to LHDI onset, the electromagnetic LHDI can cause
turbulent reconnection at multiple x-lines
before reconnection due to the initially applied
$B$-field perturbation takes place. The turbulence in the electron diffusion
region is also largely determined by the time that the LHDI has been
present before the reconnection process leads to a decay of the LHDI around the
x-line. We compared the LHDI development
in two-dimensional Vlasov and ten moment gradient simulations
with the one within three-dimensional reconnection using
the same initial conditions for the current sheet. The
electric field's evolution is given in Fig.~\ref{fig:rec_rate_lhdi_development}
(middle). Comparing the ten moment simulation of pure LHDI
with the reconnection simulation, the development is very
similar indicating that the LHDI is the main source of
fluctuations in $E_z$ as expected. Its onset is slightly
later within reconnection because of the broader
current sheet that develops in the course of the
reconnection process. Looking at the Vlasov simulation,
the development is again comparable but onset time is much
earlier and growth rate is also higher ($1.08\,\Omega_{i,0}$
compared to $0.61\,\Omega_{i,0}$ in the fluid model). The
faster onset of the LHDI using the kinetic
model in this setup is consistent with the faster onset
of reconnection so that in the end the effect of
the LHDI in ten moment gradient simulations resembles
the effect in kinetic simulations.
It should however be noted that the
initially applied perturbation of $\delta n = 10^{-8} n_0$
is small compared to intrinsic PIC noise present
in kinetic simulations of 3D reconnection.
A low initial noise level is necessary for agreement
with previous PIC studies which is interesting because
reconnection onset and LHDI onset are delayed to
similar extent compared to Vlasov simulations. The
reason is that the electromagnetic branch of the
LHDI has a high saturation level so that the magnetic field
fluctuations can initiate turbulent reconnection at multiple
x-lines.

The onset duration of the LHDI is
important because it influences the the time when
strong fluctuations in the magnetic field arise and largely
determines the impact that the LHDI can have around the x-line.
In order to examine the dependence of LHDI onset on the
noise level, we apply initial perturbations
of different magnitudes $\delta n$ to electron and ion
density in the current sheet setup used for the Harris
sheet reconnection simulations, and measure the onset duration
(time until $||E_z||_{\infty} \geq 0.04\,B_0 v_{A,0}$).
The results are shown in Fig.~\ref{fig:rec_rate_lhdi_development}
(lower). The onset duration has
an approximately logarithmic dependence $t_{onset} \sim \log(\delta n / n_0)$
in this setup using a resolution as in the three-dimensional simulation.
There is a wide range between $t_{onset} = 30\,\Omega_{i,0}^{-1}$ when
$\delta n = 10^{-10}\,n_0$ and $t_{onset} = 4\,\Omega_{i,0}^{-1}$
when $\delta n = 10^{-2}\,n_0$ and the time development is substantially
different even when comparing, for example, an initialisation
using $\delta n = 10^{-6}\,n_0$ and one using $\delta n = 10^{-4}\,n_0$.
There is also competition between the LHDI onset
and the reconnection onset, which is not significantly influenced by
the level of initial random noise but can be controlled by the magnitude
of perturbation in the magnetic flux $\psi$. Thus, depending on the
initial perturbation one can get highly turbulent reconnection with x-lines 
due to current sheet instabilities or weak effect of instabilities
and a reconnection that is essentially two-dimensional. 
In fluid simulations there is the additional
dificulty that onset times are affected by the closure approximation.
Past studies of 3D reconnection that used kinetic paricle-in-cell simulations
typically relied on the discrete particle noise to trigger instabilities.
When comparing such simulations it is important to consider that
instability onset depends on the respectively used
numbers of particles, resolutions and noise reduction techniques.
Discrepancies in past publications concerning the influence of
instabilities and turbulence may partly be attributed to varying
onset times caused by different noise levels.
Generally, a prediction of the influence of instabilities in magnetospheric
reconnection events using simulations with a pre-formed current sheet is
a difficult task because accurate information
on perturbation levels at the beginning of the simulated time span would
be needed. Three-dimensional Harris sheet
simulations provide valuable insight in the possible scenarios
concerning turbulence in magnetospheric reconnection though.

\section{Conclusions}\label{sec:discussion_summary}

The ten moment fluid model with a closure based on pressure gradients can
adequately represent magnetic reconnection and the lower-hybrid drift instability
(LHDI), making it a suitable candidate for magnetospheric modelling.
The closure is applicable in a broad range of plasma configurations
in the collisionless regime without the need for parameter tuning.
Important features of the LHDI are captured like the electrostatic
and electromagnetic fluctuations and current sheet kinking due to
the electromagnetic branch. Onset times, growth rates and saturation
levels often match kinetic results, but depending on the plasma parameters
there can be discrepancies. The full range of LHDI modes is
not accurately modelled at the same time and the electrostatic modes
can be captured more easily than the electromagnetic ones.
The fluid model is particularly useful when decently resolved
kinetic particle-in-cell (PIC) simulations
are not feasible since the computational cost of fluid simulations
is typically much lower.

In both fluid and Vlasov continuum simulations saturation levels
of the LHDI were higher than expected which enabled the
electromagnetic branch to cause
heavy turbulence. Therefore, the initial perturbation level
needed to be very low in order to match results from previous kinetic
PIC simulations of three-dimensional reconnection.
A detailed analysis of LHDI saturation levels and
potential implications will be necessary in the future.
Generally, the duration of LHDI onset (and thus its
influence on the reconnection process) is very sensitive to
perturbation levels which must be considered in the interpretation
of reconnection simulations.
In three-dimensional fluid reconnection, reconnection onset was faster
than in two dimensions and the current sheet broadened in the later
stages. Strong turbulence in the outflow and moderate turbulence in the
electron diffusion region was observed to develop.

Future work on the heat flux closure may take into account the direction
imposed by the magnetic field, which 
has not been considered at this point. The
formation of magnetic islands in connection with the ten moment gradient
model needs further investigation. Improvements to the representation
of Harris sheet reconnection onset and LHDI onset/growth rate independently
of plasma parameters are desirable.
This paper focused on the applicability of the fluid model in three-dimensional
reconnection and therefore on the turbulence generation of the LHDI. Other
effects like the role of anomalous dissipation in fluid simulations
deserve a closer look. Instabilities at later times
such as the current sheet shear instability (CSSI) at the x-line or the effect of
instabilities in the outflow regions will be addressed in the future.
The process of reconnection onset and the
connection with instabilities is not well understood and three-dimensional
simulations of reconnection that take current sheet formation
into account may provide new insights in this direction.\\

\begin{acknowledgments}

\noindent\textbf{Acknowledgements}

We gratefully acknowledge the Gauss Centre for Supercomputing e.V.
(www.gauss-centre.eu) for funding this project by providing computing time
through the John von Neumann Institute for Computing (NIC) on the GCS
Supercomputer JUWELS at Jülich Supercomputing Centre (JSC).
Computations were conducted on JUWELS \citep{juwels} and on the DaVinci cluster
at TP1 Plasma Research Department.
F.A. and S.L. acknowledge support from the Helmholtz Association (VH-NG-1239).\\

\end{acknowledgments}

\bibliographystyle{jpp}

\bibliography{references}

\end{document}